\documentclass[useAMS]{mn2e}

\usepackage{graphicx,color,soul}
\usepackage{latexsym}
\usepackage{multirow}
\usepackage{amsmath,amssymb}  
\usepackage[draft=false]{hyperref}

\title[Horizon growth of supermassive black hole seeds fed with collisional dark matter]{Horizon growth of supermassive black hole seeds fed with collisional dark matter}

\author[F. D. Lora-Clavijo, F.S. Guzm\'an and M. Gracia-Linares]{F. D. Lora-Clavijo $^1$ \thanks{E-mail:fdlora@astro.unam.mx (FDLC)},  M. Gracia-Linares $^2$ \thanks{E-mail:mgracia@ifm.umich.mx (MGL)} and F. S. Guzm\'an $^2,^3$ \thanks{E-mail:guzman@ifm.umich.mx (FSG)}   \\ 	
        $^1$ Instituto de Astronom\'{\i}a, 
        Universidad Nacional Aut\'{o}noma de M\'{e}xico, 
        AP 70-264, Distrito Federal 04510, M\'{e}xico. \\  
        $^2$ Instituto de F\'{\i}sica y Matem\'{a}ticas, Universidad
        Michoacana de San Nicol\'as de Hidalgo. \\ Edificio C3, Cd.
        Universitaria, 58040 Morelia, Michoac\'{a}n,
        M\'{e}xico.
         \\ 
        $^3$ Department of Physics and Astronomy,
        University of British Columbia. 
        6224 Agricultural Road, Vancouver BC, Canada, V6T1Z1.
        }

\begin{document}

% --->   DATE

\date{\today}

\pagerange{\pageref{firstpage}--\pageref{lastpage}} \pubyear{2014}

\maketitle

\label{firstpage}

% -----> ABSTRACT

\begin{abstract}

We present the accretion of collisional dark matter on a supermassive 
black hole seed. The analysis is based on the numerical 
solution of the fully coupled system of Einstein-Euler equations for 
spherically symmetric flow, where the dark matter is modeled as a perfect 
fluid that obeys an ideal gas equation of state. As the black hole actually grows, the accretion rate of dark matter corresponds 
to the black hole apparent horizon growth rate. 
We analyze cases with infall velocity as high as  $0.5c$ and an environment density of $100M_{\odot}/\mathrm{pc}^3$, which are 
rather extreme conditions. 
Being the radial flux the maximum accretion case, our results show that the accretion 
of an ideal gas, eventually collisional dark matter, does not contribute significantly to SMBH masses. This result favors 
models predicting SMBHs were formed already with supermasses. 
We show that despite the fact that we are solving the full general relativistic system, for the parameter space studied our results are surprisingly similar to those obtained using the Bondi formula, which somehow certifies its use as a good approximation of a fully evolving space-time with spherical symmetry at short scales at least for dark matter densities. Additionally, we  study the density profile of the gas and find that the presence of SMBHs redistributes the gas near the event horizon with a cuspy profile, whereas beyond a small fraction of a parsec it is not-cuspy anymore. 
\end{abstract}

% ----->   PACS


% ----->     INTRODUCTION     <-----

\section{Introduction}

Explaining the mass of supermassive black holes (SMBHs) is one of the most interesting problems of astrophysics. A standard approach to the problem assumes that these holes are the result of the accretion onto intermediate black hole mass seeds. The natural quest is related to the origin of such seeds, which has opened a wide set of possibilities. For instance, the SMBHs could be the results of the gravothermal collapse of dark matter cores that may collapse to form SMBHs already of observed size \cite{Balberg2002b,BalbergShapiro2002}, or it could be that disks formed in protogalaxies allow the infall of matter that collapses to form black holes of $10^5 M_{\odot}$ \cite{Bullock2004}. Seeds of $10^3 M_{\odot}$ can be formed due to the runaway collision of compact stellar clusters in low metallicity protogalaxies at $z \sim 10-20$ that additionally allow accretion during the quasar era  \cite{Devecchi}. Also seeds can be the result of the core collapse of dense clusters and form $10^5M_{\odot}$ black holes \cite{Davies2011}. or they could be the result of merger tree type of build-up \cite{Volonteri2005}.  Another alternative is that seeds of $10^5M_{\odot}$ can be formed due to the collapse of already nearly as massive stars \cite{Umeda2009}; a more detailed model in this direction proposes the supermassive primordial star forms in a region of the universe with a high molecule dissociating background radiation field, and collapses directly into $10^4 - 10^6 M_{\odot}$ seed black holes \cite{DanielHolz}. More recently, the growth of primordial black holes is also a possibility, since these can grow up to $10^3-10^6 M_{\odot}$ during the radiation dominated era \cite{LoraGuzmanCruz2013}.

On the other hand, the growth process of seeds in standard analyses of SMBH growth, based on the study of the evolution of phase space distributions, consider that SMBHs are primarily fed by collisionless dark
matter, or stars 
\cite{LighShap1977,ZhaoRees2002}. Previous results show that the timescale is too long for collisionless matter to contribute significantly to black hole
growth, for instance \cite{ReadGilmore2003}; in \cite{HernandezLee} a density threshold for massive accretion or even runaway black hole mass accretion is found, which is sufficiently higher compared to  observed dark matter densities; in \cite{Peirani2008} it is shown that dark matter contributes with at 
most $10\%$ of the total accreted mass, which together with observations of the bolometric 
quasar luminosity confirms that baryons are the main matter component that feeds the black 
hole through accretion. In \cite{GuzmanLora2011a} the conditions for stable accretion and runaway accretion have been studied for an ideal gas equation of state.  Other more general cases, considering the non-radial accretion process, require the particles to overcome the angular momentum barrier in order to get the gas to the very center of a galaxy \cite{King2006}. 

In our approach we model the dark matter as a perfect fluid that obeys an ideal gas equation of state with pressure, and study the tendency in the limit of a pressureless gas. This pressure implies that dark matter fluid elements interact not only via the gravitational field, but through self-interaction. The assumption of self-interacting dark matter (SIDM) is not new, and in fact has been used to analyze the problem of SMBH plus dark matter. For instance in \cite{Balberg2002b,BalbergShapiro2002} SIDM is used to model direct dynamical collapse and SMBH formation due to the gravothermal catastrophe, and furthermore would explain the different SMBH seed masses in terms of the redshift at which the collapse took place; in \cite{Biermann} it is shown that fermionic dark matter can feed SMBH seeds to make them grow up to $10^3 -10^6 M_{\odot}$; in \cite{Saxton} the self-interaction is introduced using a polytropic equation of state for the pressure and it is also found a mechanism for SMBH formation; in \cite {Pepe} some bounds are analyzed on the accreted mass for various equations of state including nearly stiff fluids. In \cite{Ostriker2000} it is analyzed the black hole formation due to the collapse of SIDM and also studied the SMBH formation and growth due to the collapse and accretion of SIDM, so as in \cite{Hu2006}. 

We consider it interesting to include the evolution of the space-time in order to have the formally correct black hole growth rate, calculated on a truly black hole space-time that actually grows during the process. In order to do so, we present in this paper the solution of the non-linear accretion of gas into a SMBH for the particular case of radial flows. This means that we solve the full set of Einstein equations sourced by a single gas component. Full non-linear General Relativity has been used  in spherical accretion to study stationary solutions \cite{Karkowski2006} and the stability of stationary solutions with compact support perturbations \cite{Machetal2008,Mach2009}. In this paper we also use full non-linear General Relativity for rather out of equilibrium initial conditions and track the evolution of the black hole horizon without the need of dealing with models of the distribution of particles in the phase space or the conservation equation of mass; moreover, in contrast to previous non-linear approaches, we use horizon penetrating coordinates to describe the space-time. With this we are able to track the location of the black hole horizon and at the same time allow the gas to cross the horizon. 

As a consequence of the radial accretion process, it is worth asking about the influence of the SMBH on the central dark matter density profile in galaxies. Most of the analyses of dark matter distribution are based on simulations of structure formation of cold dark matter. Among the most studied models there are the Navarro Frenk White (NFW) density profile \cite{NFW1996,NFW1997} and Moore's model \cite{Moore}. Different models provide different central density slopes of the type $\sim 1/r^{\kappa}$; in particular the
NFW and Moore profiles show different behaviors $\kappa=1$ and $\kappa=1.5$ respectively. 
Various studies indicate different slopes, $0.85< \kappa< 1.5$ depending on the conditions of the analysis of the samples from simulations, e.g. \cite{Klypin2001,Diemand2005,Stoher2006,Navarro2008}. Observations on the other hand suggest different profiles, for instance dwarf and low surface brightness galaxies are better described by a constant density core model \cite{Burkert1995,Walter2008}, specifically, the mass density profile of dwarf
galaxies shows averages of the order $\kappa \sim 0.29$ \cite{OhBlok2010}, whereas low surface brightness galaxies show $\kappa \sim 0.2$ \cite{Blok2001}. The inconsistency between analysis of simulations and theoretical halo models is called the cusp-core problem   \cite{Spergel2000} and for more recent observational evidence \cite{OhBrook,cuspcore1,cuspcore3}. In our case we raise the question of at what extent the presence of a SMBH dynamically accreting gas, affects the density profile of a core.

This question has been analyzed previously in \cite{GondoloSilk} based on a Newtonian analysis of trajectories of test particles around a compact object, later generalized to the relativistic case on actual black hole background space-times \cite{CWill2013}. In the latter case the conclusion is that the dark matter distribution  vanishes after a few Schwarzschild radii whereas near the black hole event horizon the density is rather spiky. Previous analyses involve the possibly spiky dark matter distribution around the SMBH considering the scattering of dark matter particles by surrounding stars \cite{Merrit2004}, for example in the case of Sgr $A^{*}$ it was found a cuspy profile of dark matter independent of the initial conditions within 1$\mathrm{pc}$ \cite{Gnedin2004}. It has been discussed the possibility that a considerable amount of dark matter may have fed SMBHs \cite{Zelnikov2005} and its cuspy profile near the black hole \cite{Vasiliev2008}. On the other hand, in \cite{GuzmanLora2011b} the relativistic Euler equations were solved on a fixed black hole background space-time and the density profile analysis was done for an ideal gas evolving on the fixed space-time of a  SMBH, and it was shown that the density profile is cuspy with $\kappa \lesssim 1.5$ near the black hole, whereas at distances of $0.1 \mathrm{pc}$  the profile shows $\kappa <0.3$, which is a relatively short distance that allows consistency with cored halos. Most recently, considering a general relativistic treatment, in \cite{Shapiro2014} it was found that dark matter is cuspy near the black hole with  profile $\kappa=7/4$.

In order to solve the Einstein-Euler system we use numerical methods. For the evolution of the geometry there are various possible approaches to solve Einstein equations. We use the 3+1 decomposition of the space-time and the $\mathrm{ADM}$ formulation of Einstein's equations. For the evolution of the fluid we write down Euler equations in a flux balance law fashion using the Valencia formulation \cite{valencia1997}. In our code we use geometrized units and measure time and length in units normalized with the initial mass of the black hole. In such case, a dark matter physically high density of the order of $100M_{\odot}/\mathrm{pc}^3$ is numerically extremely low, which leads to numerical inaccuracies when the calculations proceed. At this respect we decide to solve the complete system of equations using densities that allow numerical accuracy and convergence  and then extrapolate our numerical final results to the physical values of the density.

The paper is organized as follows. The whole system of equations is described in section \ref{sec:equations}. The numerical methods used to solve the evolution problem are presented in section \ref{sec:numericalmethods}. The application of our methods to the black hole plus dark matter parameters is in section \ref{sec:results}. Finally, in  \ref{sec:discussion} we summarize our  conclusions.

% -----------> SECTION    <---------------
\section{Evolution Equations} 
\label{sec:equations}

% ----->     Subsection: geometry     <-----

\subsection{Evolution of the space-time geometry}

In order to solve numerically the Einstein field equations, we use the 3+1 decomposition of space-time and the $\mathrm{ADM}$ formulation of general relativity, in which the space-time is foliated with a set of non-intersecting space-like hypersurfaces  $\Sigma_t$, see e.g. \cite{Alcubierre,Baumgarte,Rezzolla}. The space-time is described with the line element 

\begin{equation}
ds^2 = -(\alpha^2 - \beta_i \beta^i)dt^2 + 2\beta_i dx^i dt + \gamma_{ij}dx^i dx^j, \label{eq:lineelement}
\end{equation}

\noindent where $\alpha$ is the lapse function, $\beta^i$  are the shift vector components, $\gamma_{ij}$ are the components of the induced  3-metric that relates proper distances on the spatial hypersurfaces and  $x^\mu=(t,r,\theta,\phi)$ are the coordinates of the space-time. All our calculations assume geometric units $G=c=1$, that eventually will be restored when requiring physical units.

According to the $\mathrm{ADM}$ formulation of general relativity, Einstein's equations split into evolution equations for the 3-metric and the extrinsic curvature $K_{ij}$ of the hypersurfaces $\Sigma_t$ 

\begin{eqnarray}
\partial_t \gamma_{ij} &=& -2\alpha K_{ij} + \nabla_i \beta_j + \nabla_j 	\beta_i  , \\  
\partial_t K_{ij} &=& - \nabla_i \nabla_j \alpha + \alpha \left (  R_{ij} + K K_{ij} - 2K_{il}K^l_j  \right ) \\ \nonumber
&+& 4\pi\alpha \left [ (S - \rho_{\mathrm{ADM}})\gamma_{ij}  - 2S_{ij} \right]
+ \beta^l\nabla_l K_{ij} \\ \nonumber 
&+& K_{il}\nabla_j \beta^l + K_{jl}\nabla_i \beta^l,\
\end{eqnarray}  

\noindent where $\nabla_i$ denotes the covariant derivative with respect to the 3-metric, $R_{ij}$ is the Ricci tensor of the space-like hypersurfaces $\Sigma_t$ and $K=\gamma^{ij}K_{ij}$ is the trace of their  extrinsic curvature. In addition to the evolution equations, there are the Hamiltonian and momentum constraints 

\begin{eqnarray}
^{(3)}R + K^2 - K_{ij}K^{ij} - 16\pi\rho_{\mathrm{ADM}} &=& 0, \label{eq:HC}\\
\nabla_j K^{ij} - \gamma^{ij} \nabla_j K - 8\pi j^i &=& 0, 
\end{eqnarray}

\noindent where $^{(3)}R$ is the scalar of curvature associated to $\gamma_{ij}$. In these equations, the quantities $\rho_{\mathrm{ADM}}$, $j^i$, $S_{ij}$ and $S=\gamma^{ij}S_{ij}$ correspond to the local energy density, the momentum density, the spatial stress tensor and its trace respectively, measured by an Eulerian observer. These variables are obtained from the projection of the energy momentum tensor  $T_{\mu \nu}$ of matter along the space-like hypersurfaces and along the normal direction to such hypersurfaces. We stress that in this work, we will restrict to spherically symmetric black holes.

\subsection{Euler equations}
 
In order to track the evolution of a fluid coupled to the evolution of the space-time, it is necessary to write down the general relativistic Euler equations. For a generic space-time these can be derived from the local conservation of the stress-energy tensor $\nabla_{\nu} (T^{\mu \nu}) = 0$ and the local conservation of the rest mass density $\nabla_{\nu} (\rho u^{\nu}) = 0$, where $\rho$ is the proper rest mass density, $u^{\mu}$ is the four-velocity of the fluid and $\nabla_{\nu}$ is the covariant derivative consistent with the four-metric $g_{\mu \nu}$ of the space-time (\ref{eq:lineelement}). 

We assume the matter field in the above equations is that of a perfect fluid with stress-energy tensor $T_{\mu \nu} = \rho hu^{\mu} u^{\nu} + pg^{\mu \nu}$, where $p$ is the pressure, $g_{\mu \nu}$ are the components of the four-metric and $h$ the relativistic specific enthalpy given by
$h = 1 + \epsilon + p/\rho$, where $\epsilon$ is the rest frame specific internal energy density of the fluid.

It is well known that Euler equations develop discontinuities in the hydrodynamical variables even if smooth initial data are considered \cite{LeVeque}. Thus one may solve these equations using finite volume methods, as long as the system is written in a flux balance law form, which in turn requires the definition of conservative variables.

In order to obtain the general relativistic Euler equations as a set of flux balance laws, it suffices to project the local conservation equations  along the space-like hypersurfaces and the normal direction to such hypersurfaces \cite{valencia1997,Font}. A straightforward  calculation yields the set of equations in the desired form

\begin{equation}
 \partial_{t} (\sqrt{\gamma} {\bf U} ) + \partial_{i} (\sqrt{-g}{\bf F}^{i})= \sqrt{-g}{\bf S}, \label{eq:HydroEvolve}
\end{equation} 
 
\noindent where $g$ is the determinant of the four-metric (\ref{eq:lineelement}), ${\bf U}$ is a vector of conservative variables, ${\bf F}^i$ are the fluxes along each spatial direction and ${\bf S}$ is a source vector. These last quantities are given by: ${\bf U} \equiv  [D, ~ M_j, ~ \tau ]^{\mathrm{T}} = [\rho W, ~ \rho h W^2 v_j,  ~ \rho h W^2 -p - \rho W ]^{\mathrm{T}}$, ${\bf F}^i \equiv [(v^i - \beta^i/\alpha)D, ~ (v^i - \beta^i/\alpha )M_j + \delta_{j}^{i}p, ~ (v^i - \beta^i/\alpha )\tau + v^i p]^{\mathrm{T}}$ and 
${\bf S} \equiv [ 0, ~ T^{\mu \nu} g_{\nu \sigma} \Gamma^{\sigma}_{\mu j}, ~ T^{\mu 0} \partial_{\mu}\alpha - \alpha T^{\mu \nu} \Gamma^{0}_{\mu \nu}]^{\mathrm{T}}$. In these expressions,  $\gamma=\mathrm{det}(\gamma_{ij})$ is the determinant of the 3-metric, $\Gamma^{\sigma}{}_{\mu \nu}$ are the Christoffel symbols and  $v^{i}$ is the 3-velocity measured  by an Eulerian observer and defined in terms of the spatial part of the 4-velocity $u^{i}$, as $v^{i}=u^{i}/W + \beta^{i}/\alpha$, where $W$ is the Lorentz factor given by $W=1/\sqrt{1-\gamma_{ij} v^{i}v^{j}}$.

It is still necessary to close the system of equations (\ref{eq:HydroEvolve}), for which an equation of state relating $p=p(\rho,\epsilon)$ is used. We choose the gas to obey an ideal gas equation of state $p=\rho\epsilon(\Gamma -1)$, where $\Gamma$ is the adiabatic index or the ratio of specific heats. Something to stand out, is that the relativistic sound velocity $c_{s}$ for an ideal equation of state can be written as $c_{\mathrm{s}}^{2} = p\Gamma(\Gamma-1)/[p\Gamma - \rho(\Gamma -1)]$, where its asymptotic value or its maximum permitted value is $c_{\mathrm{s}_{max}}=\sqrt{\Gamma -1}$. Thus, the choice of our initial values is restricted to this condition.

Finally, the sources in the $\mathrm{ADM}$ equations in terms of the hydrodynamic variables
are: $\rho_{\mathrm{ADM}} = \rho h W^{2} - p$, $j^{i} = \rho h W^{2}v^{i}$, $S_{ij} = \rho h W^{2} v_{i} v_{j} + \gamma_{ij}p$, $S = \rho h W^{2} v_{i} v^{i} + 3p$ and are used to source the evolution of the geometry.

%% -------------- Numerical Methods 

\section{Numerical Methods}
\label{sec:numericalmethods}

We solve the Einstein-Euler system of equations for $t>0$ on a spatial domain $r \in [r_{\mathrm{exc}},r_{\mathrm{max}}]$, for a gas filling the entire domain, which is being accreted  radially into the black hole. We use Eddington-Finkelstein type of slices to describe the space-time, and therefore it is possible to choose $r_{\mathrm{exc}}$ to lie inside the black hole's event horizon, a boundary called excision, see e.g. \cite{SeidelExcision,Thornburg,GuzmanLora2012}. For a black hole with initial mass $M$, in these coordinates the event horizon is located at $r=2M$, and we choose the internal boundary inside the black hole at $r_{\mathrm{exc}}=M$ in all our runs, which defines this boundary as a space-like surface with the light cones open and pointing toward the singularity.  On the other hand, we choose the exterior boundary to be located at $r_{\mathrm{max}}=1000M$, which is sufficiently far away as to estimate an asymptotic behavior of the system. 

\subsection{Initial Data}

We set the dark matter fluid to move radially toward the black hole, with constant density initially. The value used for the density is kept as its asymptotic value, that we associate to the energy density of the cosmological environment. We also characterize the initial velocity field $v^i$ in terms of the asymptotic initial velocity $v_{\infty}$ as $v^i=(-v_{\infty}/\sqrt{\gamma_{rr}},~0,~0)$, where the relation $v^2 = v_i v^i = v_{\infty}^2$ is satisfied. Now, in order to choose the initial pressure profile, we introduce the asymptotic speed of sound $c_{s \infty}$. Once we define the value of $c_{s\infty}$ and assume the density to be initially a constant $\rho = \rho_{\mathrm{ini}}$, the pressure can be written as $p_{\mathrm{ini}} = c_{\mathrm{s} \infty}^2 \rho_{\mathrm{ini}}/(\Gamma - c_{\mathrm{s} \infty}^2 \Gamma_1)$, where $\Gamma_1=\Gamma/(\Gamma -1 )$. In order to avoid negative and zero values on the pressure, the condition $c_{\mathrm{s}\infty} < \sqrt{\Gamma - 1}$ has to be satisfied. Finally, with this value for $p_{\mathrm{ini}}$, the initial internal specific energy is reconstructed using the equation of state. Also, there is no  prescrition -as far as we can tell- about the internal energy of a gas model for dark matter, thus we chose the value of $c_{\mathrm{s}\infty}=0.08$, which, together with $\rho_{\mathrm{ini}}$, fixes the initial value of the internal energy.

Both, $v_{\infty}$ and $c_{\mathrm{s}\infty}$ are two important parameters that define the initial velocity field. We also find it useful to define the relativistic Mach number at infinity in order to parametrize our initial data ${\cal M}^R_{\infty} = W v_{\infty} /W_s c_{\mathrm{s}\infty}  = W {\cal M} /W_s $ , where $W$ is the Lorentz factor of the gas $W_s$ is the Lorentz factor calculated with the speed of sound and ${\cal M}$ is the asymptotic Newtonian Mach number, which we use to parametrize the initial configurations. When this number is bigger than one, it is said that the flow is supersonic an otherwise subsonic. 

It is worth mentioning that in the ideal case, we would solve the Hamiltonian and momentum constraints using an arbitrary gas distribution, for instance assuming a profile for the density as a source of the constraints. What is commonly done is to assume that the gas profile is localized in a bounded region, allowing the space-time to be asymptotically flat. Here we proceed in a different manner.

Since we plan to model a system that is not asymptotically flat consisting of a gas filling the entire space approaching a localized region, we switch on the evolution equations with an initially constant density profile of very low density. The constraints are not satisfied initially, however, after a few time steps the system gas plus space-time self-regulates and at a finite short time the constraints are satisfied numerically from then on. We assume that when this happens we have consistent initial data. 

The parameter space to study the non-linear accretion of SIDM onto a spherically symmetric black hole is enormous. In order to parametrize a set of initial conditions we choose the initial rest mass density of the dark matter gas that represents the dark matter environment density $\rho_{\mathrm{ini}}$, the value of $\Gamma$ that determines the nature of the gas, $v_{\infty}$ because there is no prescription a priori on what values this may have and $c_{\mathrm{s}\infty}$ or equivalently the initial internal energy $\epsilon_{\mathrm{ini}}$ of the gas or ${\cal M}$. 

\subsection{Evolution}

The domain $r\in[r_{\mathrm{exc}},r_{\mathrm{max}}]$ is discretized for both the evolution of the geometry and the gas using a base spatial resolution $\Delta r = 0.05M$ and the integration in time assumes a discretization assuming a time resolution $\Delta t = CFL \Delta r$ with $CFL=0.25$ a constant Courant Friedrichs Lewy factor.

The combined system of Einstein and Euler evolution equations is solved simultaneously in time using the method of lines, that uses a third order total variation diminishing Runge-Kutta time integrator \cite{Shu}. The right hand sides of the geometry equations are approximated using second order accurate finite differencing stencils, with the advection terms modified with the appropriate upwind stencil that causally connects the evolution between time-steps.

On the other hand, the fluid equations are discretized using a finite volume approximation together with high resolution shock capturing methods. Specifically, we use the HLLE approximate Riemann solver \cite{HLLE} in combination with the minmod linear piecewise reconstructor. The numerical fluxes and sources in (\ref{eq:HydroEvolve}) depend both on the conservative and on the primitive variables. Then, in order to express primitive variables in terms of conservative variables, we use a Newton-Raphson algorithm each time step during the evolution.

Finally, the fluid equations diverge as the density goes to zero, and in order to avoid any true vacuum we  add an artificial atmosphere in rarefied regions, that is, we define a minimum value of $\rho$ that avoids the divergence of the specific enthalpy and the errors that propagate to the other variables. We set the density of the atmosphere to $10^{-16}$, which we found to allow accuracy and consistency of our numerical results.

\subsection{Boundary Conditions}

We solve the evolution equations in a numerical domain with two artificial boundaries. On the one hand,  we implement an excision boundary inside the black hole. This can be done because we use Eddington-Finkelstein type of coordinate to describe the black hole, with a gauge control on $\alpha$ and $\beta$ that keeps ingoing null rays pointing toward the singularity with slope -1. We in fact force the gauge to satisfy the condition $\alpha/\sqrt{\gamma_{rr}}+\beta^r=1$ during the evolution in the whole spatial domain, including the excision boundary $r=r_{\mathrm{exc}}$. This property allows the gas to enter the black hole horizon, because the two null radial rays of light cones at this boundary point toward the singularity, there is no need to impose boundary conditions at this boundary.

On the other hand we also implement an artificial boundary at $r=r_{\mathrm{max}}$. At this boundary we find interesting issues. Firstly, the geometric quantities (metric functions and their derivatives) behave as waves, and usually outgoing wave boundary conditions are applied \cite{Alcubierre,Baumgarte}. Secondly, we are injecting gas to the numerical domain using an ingoing flux condition. Thus we have a situation in which we have an outgoing channel for geometry and an ingoing flux channel for matter. In order to fix this interesting situation we implemented a constraint preserving boundary condition, which consists of solving the constraints at such boundary. Specifically, we evaluate the Hamiltonian and momentum constraints at $r_{\mathrm{max}}$, then we extrapolate the values of $\gamma_{\theta\theta}$ and $K_{rr}$ from its values $r<r_{\mathrm{max}}$ to $r_{\mathrm{max}}$ and approximate the involved spatial derivatives of all the geometric variables with second order finite difference stencil. In this way the constraints are reduced to an algebraic system of two equations with two unknowns $\gamma_{rr}(r_{\mathrm{max}})$ and $K_{\theta\theta}(r_{\mathrm{max}})$. Solving the system guarantees that the Einstein's equations are satisfied at that point. The system of equations for $\gamma_{rr}$ and $K_{\theta\theta}$ at $r=r_{\mathrm{max}}$ are

\begin{eqnarray}
&&\left(\frac{2}{\gamma_{\theta\theta}}+\frac{2K_{\theta\theta}^2}{\gamma_{\theta\theta}^2}-16\pi \left(\frac{\rho h}{1-\gamma_{rr}v^rv^r}-p \right)\right)\gamma_{rr}^2\nonumber\\
&&+ \left(-\frac{2\gamma_{\theta\theta}''}{\gamma_{\theta\theta}}+\frac{\gamma_{\theta\theta}'^2}{\gamma_{\theta\theta}}+\frac{4K_{\theta\theta}K_{rr}}{\gamma_{\theta\theta}}+\frac{3\gamma_{\theta\theta}'}{\gamma_{\theta\theta}\Delta r}\right)\gamma_{rr}\nonumber\\ 
&&+\frac{\gamma_{\theta\theta}'}{\gamma_{\theta\theta}}\left(\frac{-4\hat{\gamma}_{rr}+\tilde{\gamma}_{rr}}{2\Delta r}\right) = 0, \nonumber\\%\label{eq:sistema}\\
&&K_{\theta\theta}\left(\frac{\gamma_{\theta\theta}'}{\gamma_{\theta\theta}^2} - \frac{3}{\gamma_{\theta\theta}\Delta r}\right) 
-\frac{2}{\gamma_{\theta\theta}}\left(\frac{-4\hat{K}_{\theta\theta}+\tilde{K}_{\theta\theta}}{2\Delta r}\right)\nonumber\\ 
&&- 8\pi\gamma_{rr} \left(\frac{\rho h v^r}{1-\gamma_{rr}v^r v^r}\right) - \frac{K_{rr}\gamma_{\theta\theta}'}{\gamma_{\theta\theta}\gamma_{rr}} = 0, \nonumber% \label{eq:sistema2}
\end{eqnarray}

\noindent where 
$\hat{\gamma}_{rr}=\gamma_{rr}(r_{\mathrm{max}}-\Delta r)$, 
$\tilde{\gamma}_{rr}=\gamma_{rr}(r_{\mathrm{max}}-2\Delta r)$, 
$\hat{K}_{\theta\theta}=K_{\theta\theta}(r_{\mathrm{max}}-\Delta r)$, 
$\tilde{K}_{\theta\theta}=K_{\theta\theta}(r_{\mathrm{max}}-2\Delta r)$, and all the other functions are evaluated at $r=r_{\mathrm{max}}$ and where $\Delta r$ is the resolution of the numerical domain.

\subsection{Monitoring the evolution}

{\it Apparent horizon.} We are interested in observing the black hole horizon growth. In order to do that, we track the evolution of the apparent horizon. The apparent horizon of a spherically symmetric space-time in spherical coordinates is the outermost marginally trapped surface where $\Theta = \partial_r \gamma_{\theta\theta}/\gamma_{\theta\theta}\sqrt{\gamma_{rr}} - 2K_{\theta\theta}/\gamma_{\theta\theta} =0.$, which obeys the condition of the expansion of null surfaces to be zero. This surface can be located every time step during the evolution \cite{Thornburg,GuzmanLora2012}. Then we define the radius at which this happens as the apparent horizon radius $r_{\mathrm{AH}}=\sqrt{\gamma_{\theta\theta}{}_{\mathrm{AH}}}$. Then we estimate the apparent horizon mass $M_{\mathrm{AH}}= r_{\mathrm{AH}}/2$. Since our numerical results converge with second order, consistently  with the discretization error of the numerical methods used, the final apparent horizon mass is a Richardson extrapolation of our results using the two finest resolutions in all our runs.

{\it Constraints.} In order to validate our numerical results during the evolution of the system, we monitor the accuracy and convergence of the discretized version of the Hamiltonian and momentum constraints during the evolution. In our runs we show that the Hamiltonian constraint violation converges to zero with second order.

Another important ingredient is that we tested all our tools work fine together by checking that our main results, horizon growth rate and density profile are independent of the numerical domain size. 

% ----->     RESULTS     <-----

\section{Results}
\label{sec:results}

The methods described so far are now adapted to model the radial accretion of dark matter. The properties of the system under study are those of the black hole seed and those of the gas. As described before, we restrict to the case of spherically symmetric black holes and therefore no rotation parameter is considered here. On the other hand, we model dark matter as a perfect fluid obeying an ideal gas equation of state. Even under these set of restrictions, the parameter space to explore is considerably big and involves the radial infall velocity, the adiabatic constant, the internal energy or speed of sound and the rest mass density of the gas. An additional difficulty is that there is not an a priori prescription to choose the value of the internal energy, and thus there are no particularly pointed spots of the parameter space to focus on.

\subsection{Calculation of the horizon growth rate}

The solution of the Einstein-Euler system also requires setting tractable units in the numerical code, followed by the analysis of the numerical results associated to the horizon growth rate. We proceed as follows

\begin{enumerate}

\item For given $\Gamma$, $v_{\infty}$ and $\rho$, we solve numerically the Einstein-Euler system in geometric units $G=c=1$, with the normalized initial black hole mass $M=1$, during various hundreds of units of time in units of $M$. In this way, the coordinate time and radius appear in units of $M$. Then the environment rest mass density $\rho$ is also written in these geometrized units.

\item We calculate the mass of the black hole apparent horizon during the evolution. At post processing,  we fit its growth in time with a line. This fit provides a mass growth rate of the black hole, which would be a more accurate calculation than what is usually assumed to be a mass accretion rate estimate based on approximate accretion models. Thus, we fit the black hole growth with the formula $M_{\mathrm{AH}}=A t + B$, where the fitting parameter $A$ indicates the growth rate.

\item For every combination of the parameters explored, we fit the growth rate $A$.

\item For different SMBH seed masses, we calculate how much a SMBH seed grows in time based on the values of $A$ found from our runs.

\end{enumerate}

\begin{figure}
\includegraphics[width=8cm]{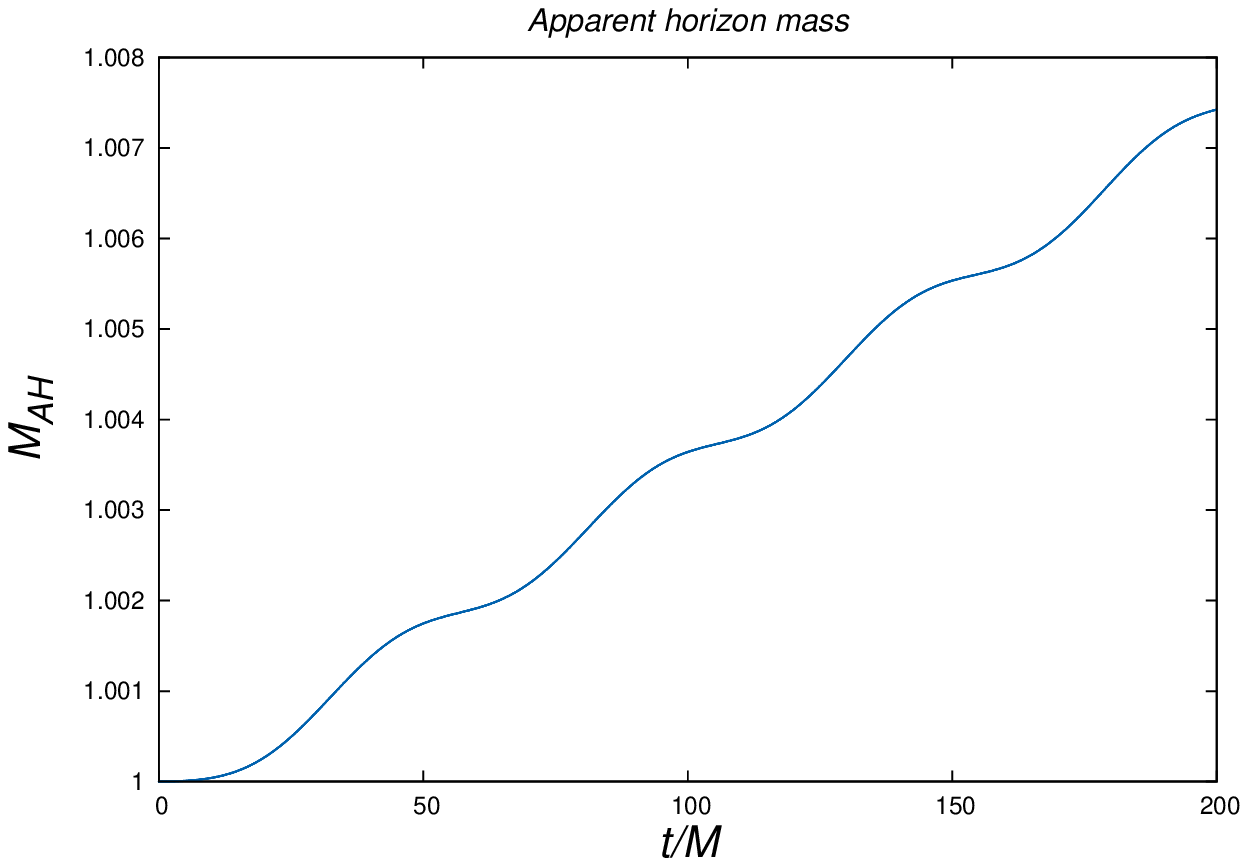}
\includegraphics[width=8cm]{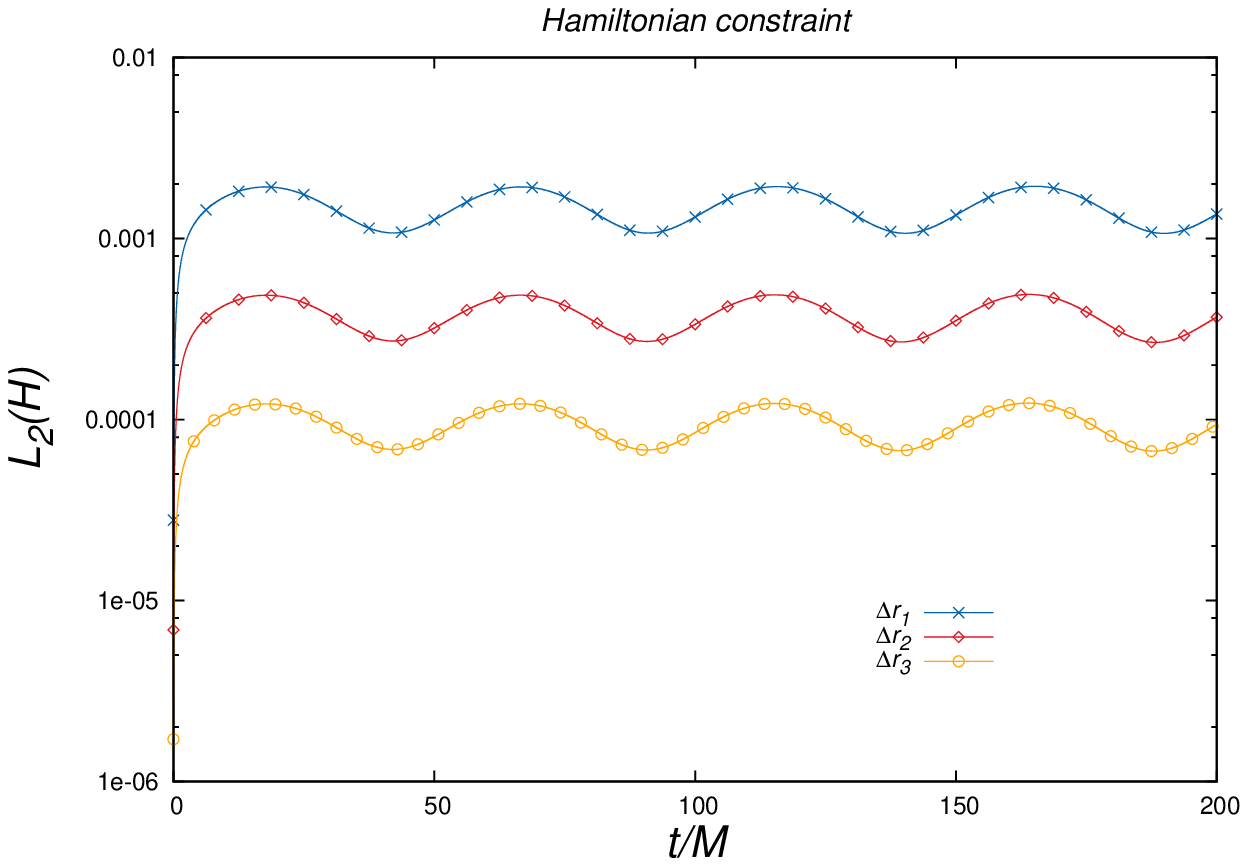}
\caption{\label{fig:example} (Top) Apparent horizon mass in time for the case $\Gamma=4/3$, $v^r=-0.1$ and $\rho=10^{-10}$. The mass horizon growth is nearly linear in time, and exemplifies the behavior of all our runs.  (Bottom) We show the $L_2$ norm of the violation of the Hamiltonian constraint. For this test we cover the domain $r\in[M,500M]$ with three different  resolutions $\Delta r_1 =0.05M $, $\Delta r_2 = \Delta r_1 /2$ and $\Delta r_3 = \Delta r_2 /2$. The error decreases quadratically with resolution, that is, the violation using resolution $\Delta r_3$ is four times smaller than when using $\Delta r_2$ and sixteen times smaller than when using $\Delta r_1$. This shows second order convergence of the violation of (\ref{eq:HC}) to zero and validates the numerical results in the continuum limit. This is the general behavior of our simulations in this paper.}
\end{figure}

In Fig. \ref{fig:example}, we illustrate for particular values of the parameters the growth of the black hole in time, and the typical behavior of all our simulations in a time window up to $200M$. The apparent horizon grows nearly linearly in time  and different growth rates are found for different values of dark matter density $\rho$ and the velocity of the gas $v^r$.  In order to validate our numerical results, one has to make sure Einstein's equations are being solved within numerical error, and for that we show also in Fig. \ref{fig:example} the $L_2$ norm of the numerical violation of the Hamiltonian constraint, calculated over the spatial domain at every time. This plot shows that the error decreases quadratically when the spatial resolution increases as expected for our numerical methods that are second order accurate as long as no shocks are formed.

With this in mind, one can observe the oscillation in the apparent horizon size, superposed to its linear growth and notice that these oscillations are correlated to the violation of the constraint and associated with numerical errors. In fact, we made sure the amplitude of the oscillations of the horizon radius decreases with resolution. Therefore the horizon growth can be considered to grow linearly. That the horizon mass growth rate depends linearly on time is the key result that allows the accretion rate estimates in this paper. The slope $A$ depends on the particular combination of the parameters $\Gamma$, the asymptotic radial velocity of the gas $v^r$ and the gas rest mass density $\rho$. The simulations use values of velocity between $v_{\infty}=0.01,~ 0.5$  and densities $\rho = 10^{-8}, 10^{-10}, 10^{-11}, 10^{-12}, 10^{-13}, 10^{-14}, 10^{-15}$, whereas the values of $\Gamma$ range from $1.01$ to $1.12$.

\begin{figure}
\includegraphics[width=8cm]{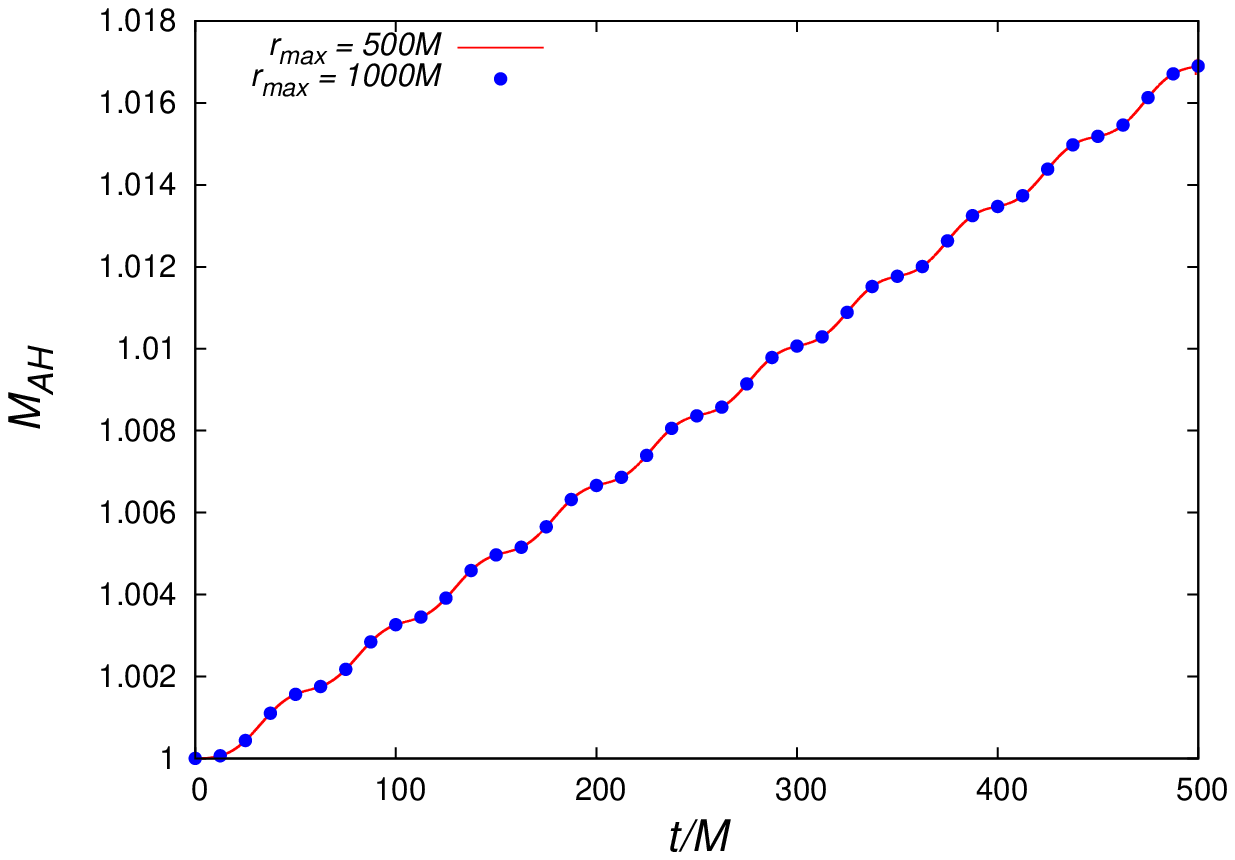}
\includegraphics[width=8cm]{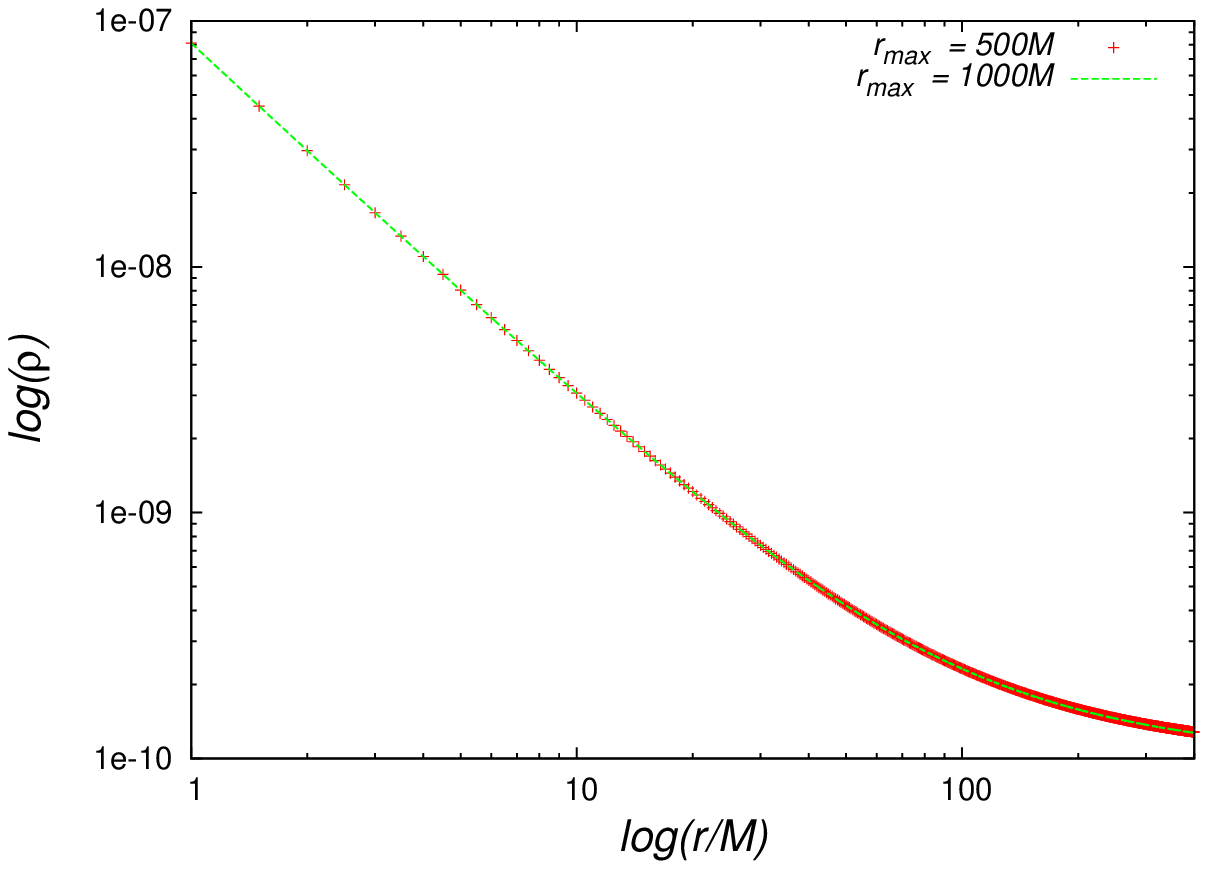}
\caption{\label{fig:BC} We show in the top panel the Apparent Horizon mass as in Fig. \ref{fig:example}, however using two different values of $r_{\mathrm{max}}=500M,~1000M$. This shows that our the results we use in our further fits are independent of the location of the external boundary. In the bottom we compare the density profile at later times suing the two numerical domains, and show it is independent of the location of the boundary as well.}
\end{figure}

Besides the accuracy and convergence of our runs, the results have to be independent of the numerical domain. In order to check this, we track the Apparent Horizon mass for two different locations of the external boundary $r_{\mathrm{max}}$ that we show in Fig. \ref{fig:BC}. The important condition we have used in all our runs is that the external boundary is located beyond the relativistic accretion radius defined by $r_{\mathrm{acc}}=\frac{M}{v^{2}_{\infty} + c_{\mathrm{s}\infty}^2}$ \cite{Petrich}, otherwise the accretion could show an unphysical runaway growth.

The accretion radius is defined in order to approximately decide when a particle does or does not fall  into the compact object, and such scale is determined by the velocity and the equation of state of the gas. Traditionally, the problem assumes the accretor is point-like, whereas in our case we are exactly doing the opposite. Numerically it seems that one has to choose between two regimes, that is, it is possible to apply the accretion of a gas to astrophysical processes if realistic velocities are considered, and consequently the accretor length scale is small enough as to be considered a point-like source that cannot be numerically resolved; conversely, if one wants to resolve the accretor (like in our case) the accretion radius length scale is unresolved unless the velocity and the sound speed of the gas are assumed to be high. This difficulty of studying both scales at the same time is the main reason why the relativistic gas accretion has not been fully resolved at the moment.

%----->

\subsection{Accretion on SMBH seeds}

In order to associate our numerical results to astrophysical scenarios, we fix the physical evolution time to 10Gyr and the environment dark matter density to $\rho=100M_{\odot} \mathrm{pc}^{-3}$. Given we use geometrized units, this density is numerically different for different values of the initial black hole seed mass. For instance, for the three seed masses of $M_{(2)}=10^3M_{\odot}$, $M_{(4)}=10^5M_{\odot}$ and $M_{(7)}=10^7M_{\odot}$, the corresponding density in geometric units is $\rho=10^{-32}$, $\rho=10^{-28}$ and $\rho=10^{-24}$ respectively. As we can see, this values of ultra low fluid density in geometrical units become numerically intractable, because their values lie below the expected round-off error threshold of our calculations. What we do is to execute runs with the numerically tractable densities $\rho = 10^{-8} - 10^{-15}$ and then extrapolate the results to the density of physically meaningful cases, which is extremely low. These extrapolated results are shown in Table {\ref{tab:extrapolating}}, where we show the result of the extrapolation of $A$ for three initial seed masses, for various values of $\Gamma$ near the $p=0$ limiting case and two values of the infall radial velocity of the gas.

\begin{table*}
\caption{\label{tab:extrapolating}Parameters fitting the apparent horizon mass growth rate for various values of the adiabatic constant $\Gamma$ and two different values of infall velocity. We show the extrapolation value corresponding to the dark matter density $\rho=100M_{\odot} \mathrm{pc}^{-3}$ which in geometric units for different black hole seed masses $M_{(2)}=10^3M_{\odot}$, $M_{(4)}=10^5M_{\odot}$ and $M_{(7)}=10^7M_{\odot}$, correspond to $\rho=1.17\times10^{-32}$, $\rho=1.17\times10^{-28}$ and $\rho=1.17\times10^{-24}$ respectively. The ansatz used to fit the mass of apparent horizon is $M_{\mathrm{AH}}(t)= At + B$. The value of $B$ for all the cases is $B=0.999$ with maximum error $0.0006\%$ as expected for an initially normalized black hole mass 
and the error in all our set of parameters is below $0.016\%$ for $A$. The values indicated under each value of $\rho$ are the values of $A$ that better fit the mass growth.}
\begin{center}
\begin{tabular}{|l|lllllll|}\hline
$\Gamma$&$v_{\infty}$& $\rho = 10^{-8}$& $\rho = 10^{-10}$ &...& $\rho = 10^{-24}$ & $\rho = 10^{-28}$ & $\rho = 10^{-32}$ \\\hline
$1.12$&$0.1$ & $3.89 \times 10^{-4}$ & $3.89 \times 10^{-6}$&...&$3.89 \times 10^{-20}$ & $3.89 \times 10^{-24}$&$3.89 \times 10^{-28}$\\
&$0.08$ & $2.63 \times 10^{-4}$&$2.63 \times 10^{-6}$ &...& $2.63 \times 10^{-20}$ & $2.63 \times 10^{-24}$ &$2.63 \times 10^{-28}$\\
$1.1$&$0.1$ & $3.86 \times 10^{-4}$ & $3.26 \times 10^{-6}$&...&$3.26 \times 10^{-20}$ & $3.26 \times 10^{-24}$&$3.26 \times 10^{-28}$\\
&$0.08$ & $2.6 \times 10^{-4}$&$2.6 \times 10^{-6}$ &...& $2.6 \times 10^{-20}$ & $2.6 \times 10^{-24}$ &$2.6 \times 10^{-28}$\\
$1.01$&$0.5$ & $4.31 \times 10^{-3}$ & $4.31 \times 10^{-5}$&...&$4.31 \times 10^{-19}$ & $4.31 \times 10^{-23}$&$4.31 \times 10^{-27}$\\
&$0.4$ & $3.9 \times 10^{-3}$ & $3.9 \times 10^{-5}$&...&$3.9 \times 10^{-19}$ & $3.9 \times 10^{-23}$&$3.9 \times 10^{-27}$\\
&$0.3$ & $3.76 \times 10^{-4}$ & $3.76 \times 10^{-6}$&...&$3.76 \times 10^{-20}$ & $3.76 \times 10^{-24}$&$3.76 \times 10^{-28}$\\
&$0.2$ & $3.46 \times 10^{-4}$ & $3.46 \times 10^{-6}$&...&$3.46 \times 10^{-20}$ & $3.46 \times 10^{-24}$&$3.46 \times 10^{-28}$\\
&$0.1$ & $3.26 \times 10^{-4}$ & $3.26 \times 10^{-6}$&...&$3.26 \times 10^{-20}$ & $3.26 \times 10^{-24}$&$3.26 \times 10^{-28}$\\
&$0.08$ & $2.98 \times 10^{-4}$&$2.98 \times 10^{-6}$ &...& $2.98 \times 10^{-20}$ & $2.98 \times 10^{-24}$ &$2.98 \times 10^{-28}$\\
&$0.06$ & $2.96 \times 10^{-4}$&$2.96 \times 10^{-6}$ &...& $2.96 \times 10^{-20}$ & $2.96 \times 10^{-24}$ &$2.96 \times 10^{-28}$\\
&$0.04$ & $2.95 \times 10^{-4}$&$2.95 \times 10^{-6}$ &...& $2.95 \times 10^{-20}$ & $2.95 \times 10^{-24}$ &$2.95 \times 10^{-28}$\\
&$0.02$ & $2.949 \times 10^{-4}$&$2.949 \times 10^{-6}$ &...& $2.949 \times 10^{-20}$ & $2.949 \times 10^{-24}$ &$2.949 \times 10^{-28}$\\
&$0.01$ & $2.941 \times 10^{-4}$&$2.941 \times 10^{-6}$ &...& $2.941 \times 10^{-20}$ & $2.941 \times 10^{-24}$ &$2.941 \times 10^{-28}$\\\hline
\end{tabular}
\end{center}
\end{table*}

\begin{table*}

\caption{\label{tab:10Gyr}Mass accreted by the a black hole during 10Gyr for six different black hole seed masses. The initial environment density of dark matter is $\rho=100M_{\odot} \mathrm{pc}^{-3}$. The seed black hole mass is indicated with $M_{(i)}$. We pay extra attention to the lowest value of $\Gamma$ and explore the subsonic and supersonic regimes.}
\begin{center}
\begin{tabular}{llllllll}\hline
$\Gamma$& $v_{\infty}$& $M_{(1)} = 10^{2}M_{\odot}$ & $M_{(2)} = 10^{3}M_{\odot}$ &$M_{(3)} = 10^{4}M_{\odot}$&$M_{(4)} = 10^{5}M_{\odot}$& $M_{(5)} = 10^{6}M_{\odot}$&$M_{(6)} = 10^{9}M_{\odot}$ \\\hline
1.12&$0.1$ & $2.49\times 10^{-9} M_{(1)}$ &$2.49\times 10^{-8} M_{(2)}$&$2.49 \times 10^{-7} M_{(3)}$& $2.49 \times 10^{-6} M_{(4)}$ & $2.49 \times 10^{-5} M_{(5)}$ & $2.49 \times 10^{-2}M_{(6)}$\\
&$0.08$ & $1.68\times 10^{-9} M_{(1)}$&$1.68\times 10^{-8} M_{(2)}$ &$1.68\times 10^{-7} M_{(3)}$&$1.68\times 10^{-6} M_{(4)}$ & $1.68\times 10^{-5} M_{(5)}$ &$1.68\times 10^{-2} M_{(6)}$\\
1.1&$0.1$ & $2.46\times 10^{-9} M_{(1)}$ &$2.46\times 10^{-8} M_{(2)}$&$2.46 \times 10^{-7} M_{(3)}$& $2.46 \times 10^{-6} M_{(4)}$ & $2.46 \times 10^{-5} M_{(5)}$ & $2.46 \times 10^{-2}M_{(6)}$\\
&$0.08$ & $1.66\times 10^{-9} M_{(1)}$&$1.66\times 10^{-8} M_{(2)}$ &$1.66\times 10^{-7} M_{(3)}$&$1.66\times 10^{-6} M_{(4)}$ & $1.66\times 10^{-5} M_{(5)}$ &$1.66\times 10^{-2} M_{(6)}$\\
1.01 &$0.5$ & $2.75\times 10^{-8} M_{(1)}$ &$2.75\times 10^{-7} M_{(2)}$&$2.75 \times 10^{-6} M_{(3)}$& $2.75 \times 10^{-5} M_{(4)}$ & $2.75 \times 10^{-4} M_{(5)}$ & $2.75 \times 10^{-1}M_{(6)}$\\
&$0.4$ & $2.5\times 10^{-8} M_{(1)}$ &$2.5\times 10^{-7} M_{(2)}$&$2.5 \times 10^{-6} M_{(3)}$& $2.5 \times 10^{-5} M_{(4)}$ & $2.5 \times 10^{-4} M_{(5)}$ & $2.5 \times 10^{-1}M_{(6)}$\\
 &$0.3$ & $2.41\times 10^{-9} M_{(1)}$ &$2.41\times 10^{-8} M_{(2)}$&$2.41 \times 10^{-7} M_{(3)}$& $2.41 \times 10^{-6} M_{(4)}$ & $2.41 \times 10^{-5} M_{(5)}$ & $2.41 \times 10^{-2}M_{(6)}$\\
 &$0.2$ & $2.21\times 10^{-9} M_{(1)}$ &$2.21\times 10^{-8} M_{(2)}$&$2.21 \times 10^{-7} M_{(3)}$& $2.21 \times 10^{-6} M_{(4)}$ & $2.21 \times 10^{-5} M_{(5)}$ & $2.21 \times 10^{-2}M_{(6)}$\\
&$0.1$ & $2.08\times 10^{-9} M_{(1)}$ &$2.08\times 10^{-8} M_{(2)}$&$2.08 \times 10^{-7} M_{(3)}$& $2.08 \times 10^{-6} M_{(4)}$ & $2.08 \times 10^{-5} M_{(5)}$ & $2.08 \times 10^{-2}M_{(6)}$\\
&$0.08$ & $1.91\times 10^{-9} M_{(1)}$&$1.91\times 10^{-8} M_{(2)}$ &$1.91\times 10^{-7} M_{(3)}$&$1.91\times 10^{-6} M_{(4)}$ & $1.91\times 10^{-5} M_{(5)}$ &$1.91\times 10^{-2} M_{(6)}$\\
&$0.06$ & $1.897\times 10^{-9} M_{(1)}$&$1.897\times 10^{-8} M_{(2)}$ &$1.897\times 10^{-7} M_{(3)}$&$1.897\times 10^{-6} M_{(4)}$ & $1.897\times 10^{-5} M_{(5)}$ &$1.897\times 10^{-2} M_{(6)}$\\
&$0.04$ & $1.891\times 10^{-9} M_{(1)}$&$1.891\times 10^{-8} M_{(2)}$ &$1.891\times 10^{-7} M_{(3)}$&$1.891\times 10^{-6} M_{(4)}$ & $1.891\times 10^{-5} M_{(5)}$ &$1.891\times 10^{-2} M_{(6)}$\\
&$0.02$ & $1.890\times 10^{-9} M_{(1)}$&$1.890\times 10^{-8} M_{(2)}$ &$1.890\times 10^{-7} M_{(3)}$&$1.890\times 10^{-6} M_{(4)}$ & $1.890\times 10^{-5} M_{(5)}$ &$1.890\times 10^{-2} M_{(6)}$\\
&$0.01$ & $1.88\times 10^{-9} M_{(1)}$&$1.88\times 10^{-8} M_{(2)}$ &$1.88\times 10^{-7} M_{(3)}$&$1.88\times 10^{-6} M_{(4)}$ & $1.88\times 10^{-5} M_{(5)}$ &$1.88\times 10^{-2} M_{(6)}$\\\hline
\multicolumn{2}{c}{Bondi} & 
$1.72\times 10^{-9}M_{(1)}$ & 
$1.72\times 10^{-8}M_{(2)}$& 
$1.72\times 10^{-7}M_{(3)}$& 
$1.72\times 10^{-6}M_{(4)}$& 
$1.72\times 10^{-5}M_{(5)}$& 
$1.72\times 10^{-2}M_{(6)}$\\\hline
\end{tabular}
\end{center}
\end{table*}

Using the extrapolated value of $A$ in Table \ref{tab:extrapolating}, we are in the position of estimating the final mass of the black hole after 10Gyr. The mass accreted by various seed black holes is shown in Table \ref{tab:10Gyr}. In all the cases, the accreted mass is small compared with the initial seed black hole mass. For the smallest seed of $100M_{\odot}$, despite the range of infall velocities used, the accreted mass after 10Gyr is of the order of $10^{-9}$ that of the initial seed mass, whereas for the most massive seed $10^9 M_{\odot}$, despite the infall velocity the mass accreted is $10^{-2}$ times the seed mass. In all the range of seed masses, the accreted mass is an amount that is considerably small compared to the seed black hole mass. 

The results in Table \ref{tab:10Gyr} are more specific for the case $\Gamma=1.01$, which is the lowest value we use. In this case we explore different values of the infall velocity ranging from subsonic with Mach 0.125 for $v_{\infty}=0.01$ until supersonic with Mach 6.25 for $v_{\infty}=0.5$. The trend is that the accreted mass is higher when the infall velocity is higher for a given value of $\Gamma$. For comparison we also show the result of applying Bondi's formula \cite{Bondi}, for the spherical accretion by a point-like accretor  $\dot{M}= \frac{4\pi G^2}{c^{3}_{\mathrm{s}}}\rho M^2$, where $M$ is the mass of the black hole seed and $\rho$ the asymptotic density of the gas. It is reassuring that this simple formula provides comparable to the fully general relativistic results we found, considering this formula does not even depend on the infall velocity at a finite radius.

Even though our work horse has been the use of $c_{\mathrm{s},\infty}$ we also performed a set of simulations for smaller values of $c_{\mathrm{s},\infty}=0.05, 0.01$ and show a pretty similar behavior and accretion rates and final accreted mass in 10Gyr.

Considering that the density of dark matter we use ($100 M_{\odot}/\mathrm{pc}^3$) is high, the infall radial velocity is also high and the process is radial, we are considering an extreme conditions scenario that should provide maximum accretion rates. Our calculations show that even under these extreme conditions, the total black hole growth due to the accretion of an ideal gas, eventually dark matter, is small.

% ----->
\subsection{Density profile}

\begin{figure}
\includegraphics[width=8cm]{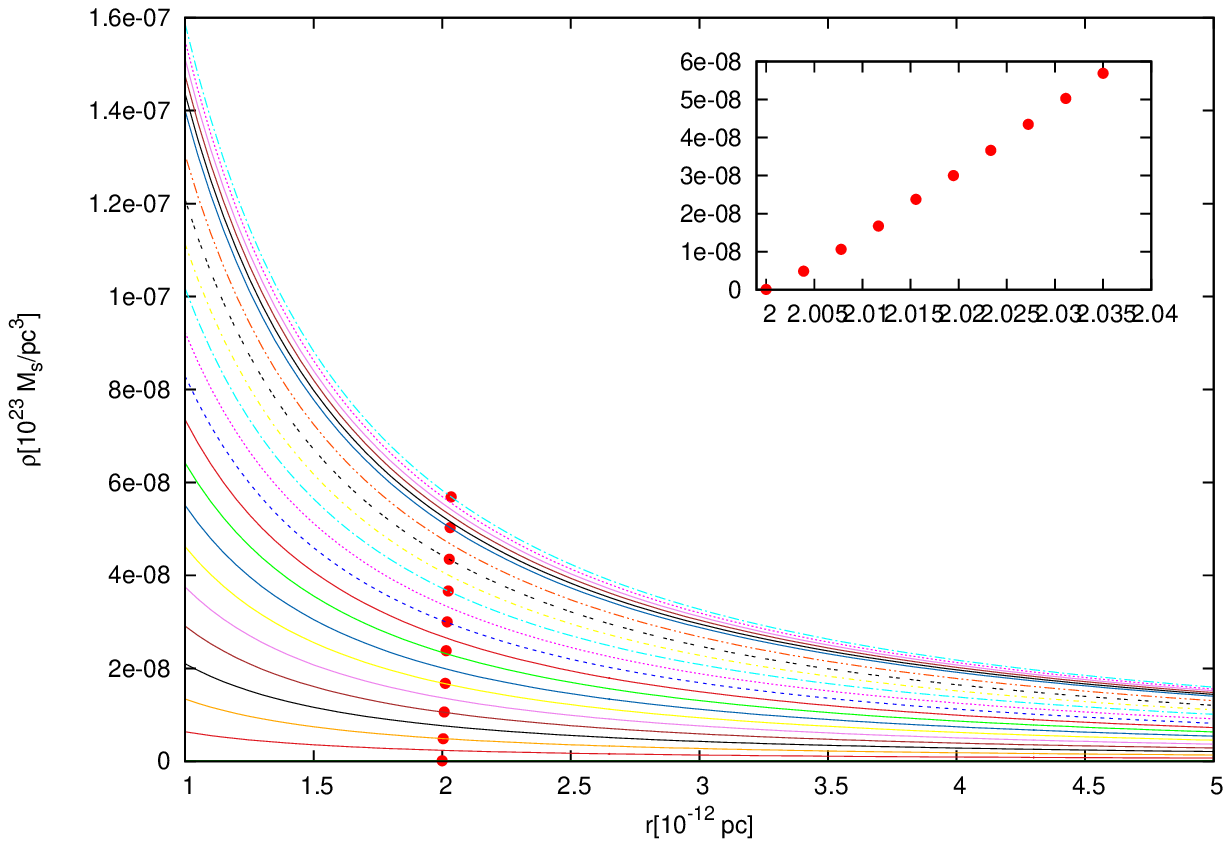}
\includegraphics[width=8cm]{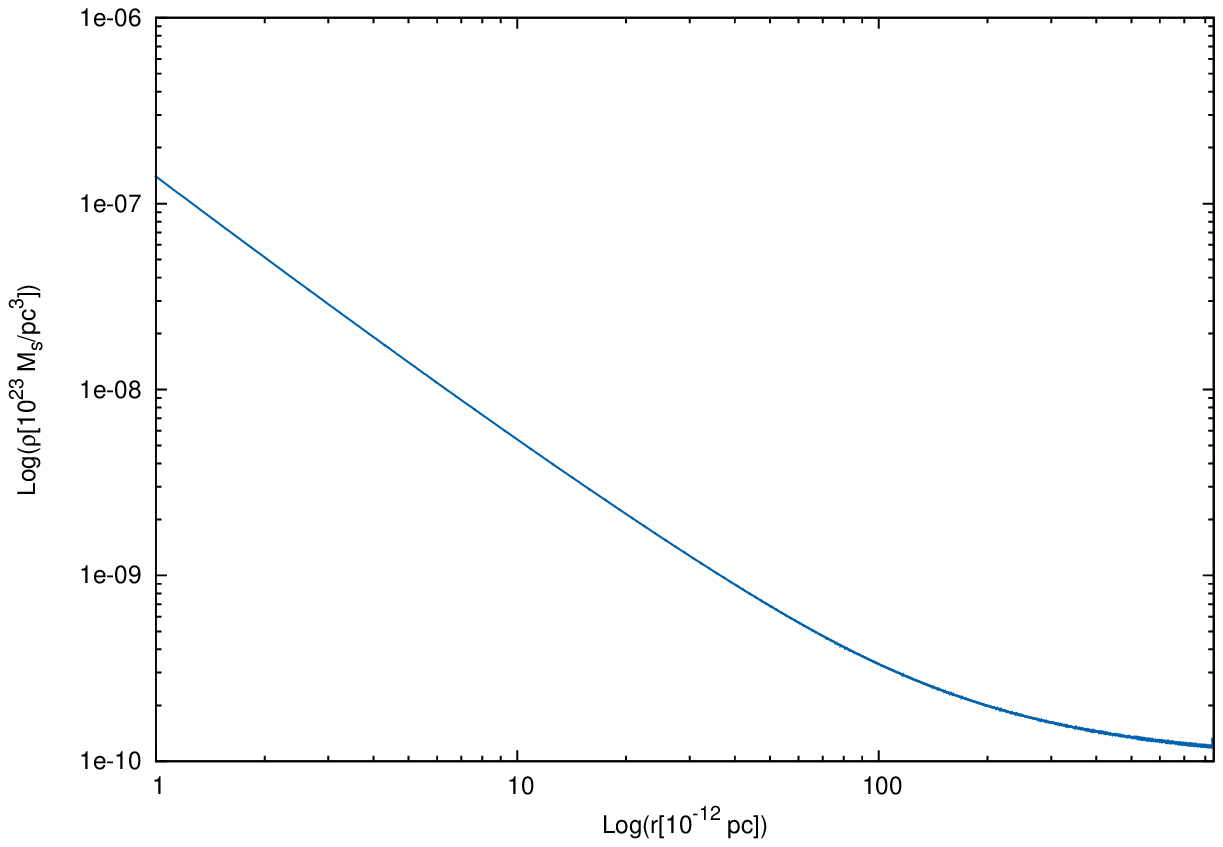}
\caption{\label{fig:profile} (Top) Snapshots of the density profile for the particular case with $\Gamma=1.1$ and  $v^r=-0.08c$ and the seed black hole mass of $10^2M_{\odot}$. Each line corresponds to a snapshot of the density, which starts being small and grows gradually until it approaches a nearly stationary stage. Lower curves correspond to earlier times and gradually lines remain nearly time-independent in the upper part. The dots indicate the location of the black hole apparent horizon radius at the time of each snapshot, and the inset shows a zoom in of its growth. The snapshots start packing  on top of each other and remain at later times. Therefore, initially the profile increases its amplitude, then stabilizes  and as the black hole grows, the profile slightly shifts outwards. (Bottom) We show that the nearly stationary profile shows a profile near the black hole that we set to be $r\in[r_{\mathrm{AH}},10M]$ and a far region for $r>500M$. For the black hole mass of $10^2M_{\odot}$ the far region starts at $500M\sim 5\times 10^{-9} \mathrm{pc}$, whereas for a grown black hole of $10^9M_{\odot}$ the far region starts at $500M\sim 5\times 10^{-2} \mathrm{pc}$.}
\end{figure}

Taking advantage of having the density profile calculated during the black hole accretion process, we ask the question of whether the dark matter density shape is cuspy around the black hole and within which scale as explored in for instance \cite{Merrit2004,Gnedin2004,Zelnikov2005,Vasiliev2008}, but in our case with the evolution of the space-time. In our simulations, the density profile starts with a constant density profile, and with the passage of time it acquires a nearly time independent sort of behavior near the horizon with a profile of type $\rho \sim 1/r^{\kappa}$. As an example, we show in Fig. \ref{fig:profile}  snapshots of the rest mass density of the gas  for one of our particular simulations. Initially the density grows near the horizon and after a finite time it accommodates around a nearly stationary profile, and for instance it does not grow out of control or runs away. The black hole horizon is growing and therefore this fact slightly modifies the density profile as shown in Fig. \ref{fig:profile}. The effect of the black hole growth shifts the density profile slightly outwards.

The fact that the density acquires a nearly stationary behavior allows one to analyze the density profile. We distinguish two regions where the nearly stationary density profile shows two different values of $\kappa$, namely $\kappa_1$ and $\kappa_2$. In all our simulations, i) a region near the black hole, where the density is cuspy up to a radius $100M$ that we parametrize with $\kappa_1$  and ii) a region far from the black hole parametrized with $\kappa_2$, that shows a flatter profile starting at about $500M$ as shown in the second panel of Fig. \ref{fig:profile} in a wide spatial domain. From the profile in the near region we can have a picture of how the gas packs around the black hole, whereas the far region actually tells us that the behavior at bigger radii, for instance at galactic core scales, where it would be interesting to know whether the black hole has an important influence and can distort the dark matter profile. We produced fits for the two regions that we show in Table \ref{tab:parameters3}.

We show that in the region near the black hole, the profile is steeper when $\Gamma$ diminishes and approaches a value $\kappa_1 \sim 1.5$ as $\Gamma \rightarrow 1$, a result very similar to that found in the test field approximation \cite{GuzmanLora2011b}.

On the other hand, in the far region we found the density profiles show a $\kappa_2<0.3$ behavior in all cases. Specifically, assuming the smallest and biggest seed mass to be $10^2M_{\odot}$ and $10^9M_{\odot}$ respectively, the numerical domain shown in Fig. \ref{fig:profile} up to 1000$M$ corresponds to $\sim 10^{-9}$ and $10^{-2} \mathrm{pc}$ respectively, which is a scale much smaller than that of halo core radii, which start at hundreds of parsecs. In this sense, SMBH seeds and SMBHs, even accreting an ideal gas with pressure radially, at high speed show a small radius of influence within a distance of a few hundreds $M$ beyond which the density profiles are not cuspy.

That the profile is steeper in the near than in the far region is expected, because the gas piles up near the black hole horizon. We stress that the fluid packs due to the non-linear coupling between geometry and fluid, and not due to the presence of an artificial boundary outside the black hole that might be reflecting material outwards, since we use horizon penetrating slices that allow the gas to enter the black hole horizon. Another important point is that the density profile is not due to the location of the external boundary as well. As an example, we show in Fig. \ref{fig:BC} how using two different numerical domains the density profile at later times is the same.

\begin{table*}
\caption{\label{tab:parameters3} We show the values of $\kappa$ assuming the $\rho(r)=C/r^{\kappa}$ density profile, for four different values of the adiabatic constant that show the tendency in the $\Gamma \rightarrow 1$ limit. We also show the results for the two separate regions, the near region $r\in[r_{\mathrm{AH}},10M]$ and the far region $r\in[500M,1000M]$. In all cases we use a rather high infall gas velocity $v^r=-0.1$ at the numerical external boundary. The results apply to two different values of black hole seed mass $10^4 M_{\odot}$ that uses an asymptotic density in geometrized units of $\rho=10^{-30}$, and $10^6 M_{\odot}$ corresponding to asymptotic density $\rho=10^{-26}$. The speed of sound used is $c_s=0.08$.}
\begin{center}
\begin{tabular}{|llll|}\hline
 & $\rho=10^{-30}$ ($10^4 M_{\odot}$)&  & $\rho=10^{-26}$ ($10^6 M_{\odot}$)\\\hline
 $\Gamma=1.12$&$\kappa_1=1.43581$& & $\kappa_1 =1.43581$  \\
 &$\kappa_2=0.21936$ & & $\kappa_2=0.21936$  \\
\hline
 $\Gamma=1.1$&$\kappa_1=1.4495$& & $\kappa_1=1.4495$  \\
 &$\kappa_2=0.239241$ & & $\kappa_2=0.239241$  \\
\hline
$\Gamma=1.05$ &$\kappa_1=1.45174$ & & $\kappa_1=1.45174$  \\
 &$\kappa_2=0.25037$ & & $\kappa_2=0.25037$  \\
\hline
$\Gamma=1.01$ &$\kappa_1=1.4525$ & & $\kappa_1=1.4525$  \\
 &$\kappa_2=0.271001$ & & $\kappa_2=0.271001$ \\
\hline
\end{tabular}
\end{center}
\end{table*}

A very important length scale to estimate the impact of these results is the relativistic accretion radius, which although is useful for stationary and very idealized processes can give a length scale idea in our rather slow accretion process. This radius is given by $r_{\mathrm{acc}}=\frac{M}{v^{2}_{\infty} + c_{\mathrm{s}\infty}^2}$ as described before \cite{Petrich}. For the value of  $c_{s,\infty}=0.08$ we used, the accretion radii corresponding to the smallest subsonic case and highest supersonic case, that is for asymptotic speed $v_{\infty}=0.01$ and $v_{\infty}=0.5$, these radii are respectively $153.8M$ and $3.9M$, which lie within what we have defined as the cuspy region near the black hole. It is good to emphasize that for values ​of $v_{\infty}$ greater than $0.5$ the accreted mass does not grow linearly anymore, on the contrary, it grows in a polynomial way or, for more extreme cases $v_{\infty}=0.9$, it grows exponentially.

These results are to be compared to a variety of different models estimating the dark matter spatial distribution profile, for instance in \cite{Vasiliev2008}, considering the scattering of stars in the bulge, it was found that solutions nearly independent of the initial conditions for the distribution of dark matter show a value $\kappa=1.5$ or bigger, whereas in simplified models calculating the influence of a central object on dark matter distribution gives $\kappa=1$ \cite{HernandezLee2008}.

% ----->     FINAL COMMENTS     <-----

\section{Discussion and Conclusions} 
 \label{sec:discussion}

We presented a fully general relativistic treatment of the radial accretion of an ideal gas onto a supermassive black hole seed, with the intention of modeling the process of accretion of collisional dark matter and show the black hole horizon growth due to accretion. Considering the density of the dark matter is ultralow, instead of using such ultralow values of the density that may propagate round-off error along the considerable amount of calculations during an evolution, we used values of the density that provide accurate and convergent calculations and at the end we extrapolate back our results to physically meaningful values.

Our results indicate that the amount of mass accreted by the black hole seed is a small portion of the initial black hole mass. The conditions assumed in our analysis are extreme, in the sense that we assume radial accretion and a considerably high infall velocity, conditions that are expected to provide a maximum accretion rate scenario. We show that the black hole horizon mass grows a small amount, for example $10^{-7}M_{\odot}$ for a seed of $100M_{\odot}$ and $10^5M_{\odot}$for a one million solar masses initial seed during 10Gyr, nearly independently of the value of $\Gamma$ we used.  We found that the formula for the Bondi accretion provides results surprisingly close to those of our non-linear calculations, even if Bondi model considers totally different hypotheses.

In this sense, our results seem to agree with models assuming SMBHs acquire their mass since very early ages, and not through accretion. That is, models predicting the black holes collapse of already supermassive size $\sim 10^6M_{\odot}$ as in \cite{Bromm}, where the collapse is modeled as the result of primordial gas clouds, or the growth of primordial black holes during the radiation dominated era, that allow the growth up to SMBH masses of up to $10^6M_{\odot}$ \cite{LoraGuzmanCruz2013}. On the other hand, the seeds of intermediate masses like the collapse of massive Pop III seeds of the order of 100$M_{\odot}$ \cite{Bromm} or those due to the runaway collisions in dense stellar clusters at $z\sim 10-20$ of $\sim 10^3 M_{\odot}$ \cite{Madau}, assumed to grow later on due to accretion processes would be inconsistent with our calculations.

Finally, concerning the density profiles, what we found is that assuming a profile $1/r^{\kappa}$, near the black hole $\kappa=\kappa_1<1.5$ in all cases analyzed, whereas in the far region $\kappa=\kappa_2<0.3$. By far region we set the limit starting at a radius of $500M$ from the origin, which for the most massive black hole considered here of $10^9 M_{\odot}$ corresponds to a distance of the order of $5\times  10^{-2} \mathrm{pc}$ from the black hole. This means that the presence of the black hole distorts the matter distribution in a cuspy way only within such distance. Despite the fact that we are injecting matter permanently in the radial direction, that is, the highest accretion rate possible, we conclude that the black hole does not produce any noticeable spiky a dark matter distribution at distances of already a small fraction of a parsec, and therefore the SMBH is not a problem to non-cuspy halo models.

% ----->     ACKNOWLEDGMENTS     <-----

\section*{Acknowledgments}

This research is partly supported by grants: 
CIC-UMSNH-4.9 and CONACyT 106466. We appreciate the comments from the anonymous referee.
%F.S.G. acknowledges support from the CONACyT program for sabbatical visits in foreign countries and the kind hospitality of UBC.

% -------------------------------------------------------
% -----     REFERENCES     ----------
% -------------------------------------------------------

\bsp

\label{lastpage}


\begin{thebibliography}{99}

\bibitem[Alcubierre 2008]{Alcubierre} Alcubierre M. 2008,
	             {\it Introduction to 3+1 Numerical Relativity}.
	             Oxford Science Publications, Oxford).

\bibitem[Amorisco et al. 2013]{cuspcore3}
	Amorisco, N. C. Agnello, A. and Evans, N. W. 2013,
	MNRAS {\bf 429}, L89-L93, arXiv:1210.3157 [astro-ph.CO].


\bibitem[Balberg et al. 2002]{Balberg2002b}
	Balberg, S., Shapiro, S. L. and Inagaki, S. 2002.
	ApJ {\bf 568} 475-487, arXiv:astro-ph/0110561.
	
\bibitem[Balberg \& Shapiro 2002]{BalbergShapiro2002}
	Balberg, S. and Shapiro, S. L. 2002,
	Phys. Rev. Lett.  {\bf 88} 101301, arXiv:astro-ph/0111176.

\bibitem[Bondi 1952]{Bondi}
        Bondi, H. 1952, MNRAS {\bf 112}, 195.

\bibitem[Banyuls et al. 1997]{valencia1997}
        Banyuls, F. et al. 1997, 
        ApJ {\bf 476}, 221-231.

\bibitem[Baumgarte 2010]{Baumgarte} Baumgarte T.W. and Shapiro, S.L. 2010,
	            {\it Numerical Relativity: Solving Einstein's equations on the Computer}.
	            Cambridge University Press, Cambridge.


\bibitem[de Blok et al. 2001]{Blok2001}
        de Blok, W.J.G, McGaugh, S. S., Bosma, A. and Rubin, V. C. 2001,
        ApJ {\bf 552}, L23-L26, arXiv:astro-ph/0103102.

\bibitem [Bromm \& Loeb 2003]{Bromm}
	Bromm, V. and Loeb, A. 2003,  
        ApJ {\bf 596}, 34-46, arXiv:astro-ph/0212400.
        

\bibitem[Burkert 1995]{Burkert1995}
        Burkert, A. 1995,
        ApJL {\bf 447}, L25, arXiv:astro-ph/9504041.

\bibitem[Davies \& Laor 2011]{Davies2011}
	Davies, S. W. and Laor, A. 2011,
	ApJ {\bf 728}, 98, arXiv:1012.3213 [astro-ph.CO].


\bibitem[Devecchi \& Volonteri 2009]{Devecchi}
	Devecchi B. and Volonteri, M. 2009, 
	ApJ {\bf 694}, 302-313, arXiv:0810.1057 [astro-ph].

\bibitem[Diemand et al. 2005]{Diemand2005}
        Diemand, J., Zemp, M., Moore, B., Stadel, J. and Carollo, C. M. 2005,
        MNRAS, {\bf 364}, 665, arXiv:astro-ph/0504215.

\bibitem[Font et al. 2000]{Font}
        Font, J. A., Miller, M., Suen, W-M. and Tobias M. 2000, Phys. Rev.
        D {\bf 61}, 044011, arXiv:gr-qc/9811015.

\bibitem[Gondolo \& Silk 1999]{GondoloSilk}
	Gondolo, P. and Silk, J. 1999, 
        Phys. Rev. Lett. {\bf 83}, 1719, arXiv:astro-ph/9906391. 

\bibitem[Gnedin \& Primack 2004]{Gnedin2004}
	Gnedin, O. Y. and Primack, J. R. 2004,
	Phys. Rev. Lett. {\bf 93}, 061302, arXiv:astro-ph/0308385.

\bibitem[Guzm\'an \& Lora-Clavijo 2011a]{GuzmanLora2011a}
	Guzm\'an, F. S. and Lora-Clavijo, F.D. 2011a,
	MNRAS {\bf 415}, 225-234, arXiv:1103.5497 [astro-ph.GA]. 
		
\bibitem[Guzm\'an \& Lora-Clavijo 2011b]{GuzmanLora2011b}
	Guzm\'an, F.S. and Lora-Clavijo, F.D. 2011b,
	MNRAS {\bf 416}, 3083-3088, arXiv:1106.3521 [astro-ph.GA].

\bibitem[Guzm\'an \& Lora-Clavijo 2012]{GuzmanLora2012}
	Guzm\'an, F.S. and Lora-Clavijo, F. D. 2012,
	Phys. Rev. D 85, 024036, arXiv:1201.3598 [astro-ph.CO].

\bibitem[Harten et al. 1988]{HLLE}
        Harten, A.,  Lax, P. D.,  and van Leer, B.,  1983, SIAM Rev. {\bf 25}, 35.
        Einfeldt, B. 1988, SIAM, J. Num. Anal. {\bf 25}, 294.

\bibitem[Hern\'andez \& Lee 2008]{HernandezLee2008}
	Hernandez, X. and Lee W. H. 2008,
	MNRAS {\bf 387}, 1727, arXiv:0803.1507 [astro-ph].

\bibitem[Hern\'andez \& Lee 2010]{HernandezLee}
	Hernandez, X. and Lee W. H. 2010,
	MNRAS {\bf 404}, L6-L10, arXiv:1002.0553 [astro-ph.CO].


\bibitem[Hu et al. 2006]{Hu2006}
        Hu, J., Shen, Y., Lou, Y-Q and Zhang, S. 2006,
        MNRAS {\bf 365}, 345-351, arXiv:astro-ph/0510222.


\bibitem[Johnson et al. 2013]{DanielHolz}
	Johnson, J.L, Whalen, D. J.  Li, H. and Holz, D. E. 2013,
	ApJ {\bf 771}, 116, arXiv:1211.0548 [astro-ph.CO].

\bibitem[Karkowski et al. 2006]{Karkowski2006}
	Karkowski, J., Kinasiewicz, B., Mach, P., Malec, E. and Swierczynski, Z. 2006,
	Phys. Rev. D {\bf 73}, 021503, arXiv:gr-qc/0509079.

\bibitem[e.g. King \& Pringle 2006]{King2006}
         King, A. R. and Pringle, J. E. 2006,
         MNRAS {\bf 373}, L90-L92, arXiv:astro-ph/0609598.

\bibitem[Klypin et al. 2001]{Klypin2001}
        Klypin, A., Kravtsov, A. V., Bullock, J. S. and Primack, J. R. 2001,
        ApJ {\bf 554}, 903, arXiv:astro-ph/0006343.

\bibitem[Koushiappas \& Bullock 2004]{Bullock2004}
        Koushiappas, S. M., Bullock, J. S. and Dekel, A. 2004,
        MNRAS {\bf 354}, 292, arXiv:astro-ph/0311487.

\bibitem[LeVeque 1992]{LeVeque} 
	LeVeque, R. J. 1992,
        {\it Numerical Methods for Conservation Laws,
        Lectures in Mathematics}, Birkh\"asuer.

\bibitem[e.g. Lightman \& Shapiro 1977]{LighShap1977}
         Lightman, A. P. and Shapiro, S. L. 1977,
         ApJ {\bf 211}, 244-262.


\bibitem[Lora-Clavijo et al. 2013]{LoraGuzmanCruz2013}
	Lora-Clavijo, F.D. Guzm\'am, F. S. and Cruz-Osorio, A. 2013,
	JCAP {\bf 12}, 015, arXiv:1312.0989 [astro-ph.CO].
	
\bibitem[Mach 2009]{Mach2009} 
	Mach, P. 2009,
	Rep. Math. Phys. {\bf 64}, 257-269, 
	
\bibitem[Mach \& Edward 2008]{Machetal2008} 
	Mach, P. and Malec, E. 2008
	Phys. Rev. {\bf D 78}, 124016, arXiv:0812.1762 [gr-qc].

\bibitem[Madau \& Rees 2001]{Madau}
	Madau, P. and Rees, M. J. 2001, 
        ApJL, {\bf 551}, L27, arXiv:astro-ph/0101223.

\bibitem[Memola et al. 2011]{cuspcore1}
	Memola, E., Salucci, P. and  Babic, A. 2011,
	A \& A, {\bf 534}, 50, arXiv:1106.5133 [astro-ph.CO].

\bibitem[Merrit 2004]{Merrit2004}
	Merritt, D. 2004,
	Phys. Rev. Lett. {\bf 92}, 201304, arXiv:astro-ph/0311594.

\bibitem[Moore et al. 1999]{Moore}
        Moore, B., Quinn, T., Governato, F. and Lake, G. 1999,
        MNRAS {\bf 310}, 1147, arXiv:astro-ph/9903164.

\bibitem[Munyaneza \& Biermann 2005]{Biermann}
	Munyaneza, F. and Biermann, P. L. 2005,
	A \& A, {\bf 436}, 805-815, arXiv:astro-ph/0403511.

\bibitem[Navarro et al. 1996]{NFW1996}
        Navarro, J. F., Frenk, C. S. and White, S. D. M. 1996,
        ApJ {\bf 462}, 563, arXiv:astro-ph/9508025.

\bibitem[Navarro et al. 1997]{NFW1997}
        Navarro, J. F., Frenk, C. S. and White, S. D. M. 1997,
        ApJ {\bf 490}, 493, arXiv:astro-ph/9611107.

\bibitem[Navarro et al. 2008]{Navarro2008}
        Navarro, J. F. et al.  2010,
        MNRAS {\bf 402}, 21-34, arXiv:0810.1522 [astro-ph].

\bibitem[Oh et al. 2010]{OhBlok2010}
        Oh, S-H., de Blok W. J. G., Brinks, E., Walter, F. and Kennicutt, R. C. 2010,
        AJ  {\bf 141}, 193, arXiv:1011.0899 [astro-ph.CO].

\bibitem[e. g. Oh et al. 2011]{OhBrook}
	Oh, S-H, Brooks, C., et al. 2011, 
	AJ {\bf 142}, 24, arXiv:1011.2777 [astro-ph.CO].
	
\bibitem[Ostriker  2000]{Ostriker2000}
         Ostriker, J. P. 2000,
         Phys. Rev. Lett, {\bf 84}, 5258-5260, arXiv:astro-ph/9912548.

\bibitem[Papadopoulos \& Font  2000]{FontPapa}
	Papadopoulos, P. and Font, J. A. 2000,
	Phys.Rev. {\bf D 61}, 024015, arXiv:gr-qc/9912054.

\bibitem[Peirani \& Freitas 2008]{Peirani2008}
        Peirani, S. and de Freitas-Pacheco, J. A., 2010, 
        Phys. Rev. {\bf D 77}, 064023, arXiv:0802.2041 [astro-ph].

\bibitem[Pepe, Pellizza \& Romero 2012]{Pepe}
	Pepe, C. Pellizza, L. J. and Romero, G. E. 2012,
	MNRAS {\bf 420}, 3298-3302, arXiv:1111.5605 [astro-ph.HE].

\bibitem[Petrich 1989]{Petrich} 
	Petrich, L. I., Shapiro, S. L., Stark, R. F. and Teulkolsky, S. A. 1989, 
	ApJ {\bf 336}, 313.

\bibitem[Read \& Gilmore  2003]{ReadGilmore2003}
        Read, J.I. and Gilmore, G. 2003,
        MNRAS {\bf 339}, 949-956, arXiv:astro-ph/0210658.

\bibitem[Rezzolla \& Zanotti 2013]{Rezzolla} 
	Rezzolla, L. and Zanotti, O. 2013,
	{\it Relativistic Hydrodynamics}. Oxford University Press, Oxford.

\bibitem[Sadeghian et al. 2013]{CWill2013}
	Sadeghian, L., Ferrer, F. and Will, C. M. 2013, 
        Phys. Rev. {\bf D 88}, 063522, arXiv:1305.2619 [astro-ph.GA].

\bibitem[Saxton \& Kinwah 2008]{Saxton}
	Saxton, C. J. and Wu, K. 2008,
	MNRAS {\bf 391}, 1403-1436, arXiv:0809.3795 [astro-ph]. 

\bibitem[Seidel \& Suen 1992]{SeidelExcision}
	Seidel, E. and Suen, W-M. 1992,
        Phys. Rev. Lett. {\bf 69}, 1845, arXiv:gr-qc/9210016.

\bibitem[Shapiro \& Paschalidis 2014]{Shapiro2014}
	Shapiro S. L. and Paschalidis, V. 2014,
	Phys. Rev. {\bf D 89}, 023506, arXiv:1402.0005 [astro-ph.CO].

\bibitem[Shu \& Osher  1989]{Shu} 
	Shu, C.W. and Osher, S.J. 1989, 
	J. Comp. Phys. {\bf 83}, 32 
	
\bibitem[Spergel \& Steinhardt  2000]{Spergel2000}
        Spergel, D. N. and Steinhardt, P. J. 2000,
        Phys. Rev. Lett. {\bf 84}, 3760-3763, arXiv:astro-ph/9909386.

\bibitem[Stoher 2006]{Stoher2006}
        Stoher, F. 2006,
        MNRAS {\bf 365}, 147.

\bibitem[Thornburg  1999]{Thornburg}
	Thornburg, J. 1999,
        Phys. Rev.  {\bf D 59}, 104007, arXiv:gr-qc/9801087.

\bibitem[Umeda et al.  2009]{Umeda2009}
        Umeda, H., Yoshida, N. et al. 2009,
        JCAP {\bf 8}, 24, arXiv:0908.0573 [astro-ph.CO].

\bibitem[Vasiliev \& Zelnikov 2008]{Vasiliev2008}
	Vasiliev, E. and Zelnikov, M. 2008,
	Phys. Rev. {\bf D 78}, 083506, arXiv:0803.0002 [astro-ph].

\bibitem[e.g. Volonteri \& Rees  2005]{Volonteri2005}
         Volonteri, M. and Rees, M. J. 2005,
         ApJ {\bf 633}, 624-629, arXiv:astro-ph/0506040.

\bibitem[Walter et al.  2008]{Walter2008}
        Walter, F., Brinks, E. et al. 2008,
        AJ {\bf 136}, 2563, arXiv:0810.2125 [astro-ph].

\bibitem[Zelnikov \& Vasiliev 2005]{Zelnikov2005}
	Zelnikov, M. I. and Vasiliev, E. A. 2005,
	Int.J.Mod.Phys. {\bf A 20}, 4217, arXiv:astro-ph/0307524.

\bibitem[Zhao et al. 2002]{ZhaoRees2002}
         Zhao, H., Haehnelt, M. G. and Rees, M. J. 2002, 
         New Astronomy, {\bf 7}, 385-394, arXiv:astro-ph/0112096.


\end{thebibliography}
\end{document}